\documentclass{article}
\usepackage{bm,latexsym,amsmath,amssymb,amsfonts,fancyhdr,color,graphicx,multirow,cite}
\usepackage[a4paper,bottom=2cm,top=2.5cm,head=5mm,width=18cm,dvipdfm]{geometry}
\usepackage[usenames,dvipsnames,svgnames,table]{xcolor}
\usepackage[colorlinks=true,
            linkcolor=blue,
            urlcolor=blue,
            citecolor=green,          
						bookmarks=true,
						bookmarksnumbered=true,
						breaklinks=true,
						pdfpagemode=Fullscreen,
						pdfstartview=FitBH]{hyperref}
\usepackage[dotinlabels]{titletoc}
\usepackage{titlesec}
\usepackage{authblk}

\numberwithin{equation}{section}
\allowdisplaybreaks[4]

\titlelabel{\thetitle.\quad \hspace{-0.8em}}
\titlecontents{section}
              [1.5em]
              {\vspace{4mm} \large \bf}
              {\contentslabel{1em}}
              {\hspace*{-1em}}
              {\titlerule*[.5pc]{.}\contentspage}
\titlecontents{subsection}
              [3.5em]
              {\vspace{2mm}}
              {\contentslabel{1.8em}}
              {\hspace*{.3em}}
              {\titlerule*[.5pc]{.}\contentspage}
\titlecontents{subsubsection}
              [5.5em]
              {\vspace{2mm}}
              {\contentslabel{2.5em}}
              {\hspace*{.3em}}
              {\titlerule*[.5pc]{.}\contentspage}

\newcommand{\titledef}{Large Leptonic Dirac CP Phase from Broken Democracy with Random Perturbations}

\hypersetup{ pdfauthor = {Shao-Feng Ge},
	     pdftitle = {\titledef}, 
	     pdfsubject = {}, 
             pdfkeywords = {}, 
	     pdfcreator = {LaTeX with hyperref package},
	     pdfproducer = {dvips + ps2pdf} }

\definecolor{gesfpurple}{rgb}{0.47,0.19,0.42}

\definecolor{gesflanse}{rgb}{0.00,0.50,0.50}

\definecolor{gesfblue}{rgb}{0.08,0.42,0.76}

\definecolor{gesfred}{rgb}{1,0,0}

\definecolor{gesfwhite}{rgb}{1,1,1}

\definecolor{gesfblack}{rgb}{0,0,0}

\newcommand{\gsec}[1]{{\hypersetup{linkcolor=red}Sec.~\ref{#1}\hypersetup{linkcolor=blue}}}

\newcommand{\geqn}[1]{\hypersetup{linkcolor=blue}(\ref{#1})\hypersetup{linkcolor=blue}}
\newcommand{\gfig}[1]{{\hypersetup{linkcolor=violet}Fig.~\ref{#1}\hypersetup{linkcolor=blue}}}

\graphicspath{{figs/}}

\begin{document}
\fontsize{12pt}{14pt}\selectfont

\title{
       \textbf{\fontsize{16pt}{18pt}\selectfont \titledef}}
\author[1,2]{{\large Shao-Feng Ge} \footnote{\href{mailto:gesf02@gmail.com}{gesf02@gmail.com}}}
\author[1,3]{{\large Alexander Kusenko} \footnote{\href{mailto:kusenko@ucla.edu}{kusenko@ucla.edu}}}
\author[1,4]{{\large Tsutomu T. Yanagida} \footnote{\href{mailto:tsutomu.tyanagida@ipmu.jp}{tsutomu.tyanagida@ipmu.jp}}}
\affil[1]{\small Kavli IPMU (WPI), UTIAS, The University of Tokyo, Kashiwa, Chiba 277-8583, Japan}
\affil[2]{\small Department of Physics, University of California, Berkeley, CA 94720, USA}
\affil[3]{\small Department of Physics and Astronomy, University of California, Los Angeles, CA 90095-1547, USA}
\affil[4]{\small Hamamatsu Professor}
\date{\today}

\maketitle

\vspace{-100mm}
\hfill IPMU18-0043
\vspace{95mm}

\begin{abstract}
A large value of the leptonic Dirac CP phase can arise
from broken democracy, where the mass matrices are democratic up to small random perturbations.  Such perturbations are a natural consequence of broken residual  $\mathbb S_3$ symmetries that dictate the democratic mass matrices at
leading order. With random perturbations, the leptonic Dirac CP phase has a higher probability to attain a value around $\pm \pi/2$. Comparing with the anarchy model,
broken democracy can benefit from residual $\mathbb S_3$
symmetries, and it can produce much better, realistic
predictions for the mass hierarchy, mixing angles, and Dirac CP 
phase in both quark and lepton sectors. Our approach provides
a general framework for a class of models in which a residual symmetry
determines the general features at leading order, and where, in the absence of other fundamental principles, the symmetry breaking appears in the form of random perturbations. 
\end{abstract}

\section{Introduction}

Flavor mixing has been observed in both quark and lepton sectors.
Theoretically, there are two basic approaches in explaining the mixing patterns.
One is top-down by assigning some full flavor symmetry to constrain the
fundamental Lagrangian.
However, the full flavor symmetries have to be broken, otherwise
the up- and down-type fermions would be subject to the same flavor structure
in their mass matrices and hence we obtain the same mixing matrices, $V_u = V_d$, leading to a trivial
physical mixing matrix, $V_{\rm CKM} = V^\dagger_u V_d = I$. If the mixing pattern
is really determined by symmetry, it has to be a residual symmetry that
survives symmetry breaking. The reverse of the top-down logic is
the bottom-up {\it phenomenological mass matrix approach}
\cite{Weinberg:1977hb,Fritzsch:1977za,Fritzsch:1979zq,FTY,TWY}.
By reconstructing the residual symmetries in both up-
and down-type fermion sectors \cite{Lam:2008rs, Lam:2008sh}, the full flavor symmetry can be obtained as
a product group \cite{Hernandez:2012ra, Hernandez:2012sk, Esmaili:2015pna}.

In both approaches, the residual symmetry takes the role of predicting the
mixing pattern \cite{Dicus:2010yu, Ge:2011ih, Ge:2011qn,Fonseca:2014koa}.
In some sense, the residual symmetry takes the same role as the custodial
symmetry in the gauge sector \cite{Ge:2014mpa}. The electroweak
$SU(2)_L \times U(1)$ gauge theory can predict the existence of four gauge
bosons but not the weak mixing angle $\theta_w$. In the Standard Model, the
weak mixing angle $\sin \theta_w = g' / \sqrt{g^2 + g'^2}$ is a function of
the gauge couplings $g$ and $g'$ whose values cannot be predicted by gauge symmetry.
The custodial symmetry can make correlation between physical observables,
$\cos \theta_w = M_W / M_Z$. Likewise, residual symmetries can predict the correlation among mixing angles and the Dirac CP
phase, also known as {\it sum rule} of mixing angles \cite{Agarwalla:2017wct,Petcov:2017ggy,Pasquini:2018udd} in addition to the mass sum rules \cite{Barry:2010yk, Gehrlein:2016wlc}. The custodial symmetry is essentially
a residual symmetry. In this sense, the concept of residual
symmetry and phenomenological mass matrix approach applies universally for
all the observed mixing among fundamental particles.

The mixing patterns in the quark and lepton sectors are quite different.
While the mixings in the quark sector are small, the lepton sector has
large mixing angles. This seems hard to understand at the first glance. Considering
the fact that the observed quark and lepton mixings are combined effects
of mixings in both up- and down-type fermions, $V_{\rm CKM} = V^\dagger_u V_d$
for quarks and $V_{\rm PMNS} = V^\dagger_\ell V_\nu$ for leptons, a unified picture
might be possible if a similar mixing pattern appears in both up- and
down-type quark mass matrices but in only one of leptons and neutrinos,
$V_u \approx V_d \approx V_\ell \mbox{ or } V_\nu$. Then the quark mixing
$V_{\rm CKM}$ is close to unity matrix while the neutrino mixing
$V_{\rm PMNS} \approx V_\ell \mbox{ or } V_\nu$ has large mixing angles \cite{FTY,TWY, Ge:2014mpa, Koide:2002cj, Joshipura:2005vy}.

Approximate democratic matrix is known as an interesting
possibility to explain the large hierarchy in quark masses and the small
CKM mixing angles when applied to both up- and down-type quarks.
If one applies this hypothesis to the lepton sector for both charged leptons
and neutrinos, with the help of residual $\mathbb S_3$ symmetries,
one may get small lepton mixing angles which is strongly
excluded experimentally. However, it was pointed out that if we assume
almost diagonal mass matrix for neutrinos we obtain large lepton mixing
angles \cite{FTY,TWY}. A natural consequence is that $V_{\rm CKM}$ has
only 1-2 mixing while $V_{\rm PMNS}$ can have at least two large mixing
angles at leading order, as we would elaborate in \gsec{sec:M0}.
The democratic matrix can also explain the mass hierarchies among charged
fermions, with $m_1 = m_2 = 0$. To accommodate nonzero
fermion masses and to get a better fit for observed mixing angles, one needs deviations from 
the democratic matrix which break the residual $\mathbb S_3$ symmetries. With residual $\mathbb S_3$ symmetries broken, there is no fundamental principle to regulate the deviations.
A natural approach is to assume that small random perturbations make the mass
matrix different from the democratic form. This approach is different from the anarchy model, where the mass matrix can be totally free of any constraint \cite{Hall:1999sn,Haba:2000be}.
We will elaborate our predictions for both neutrino
(see \gsec{sec:random} and \gsec{sec:comparison}) and quark
mixings (see \gsec{sec:quark}). 

\section{Democratic Mass Matrix Hypothesis -- Preliminary}
\label{sec:M0}

The democratic pattern of mass matrix can be realized by applying two
independent residual $\mathbb S^L_3$ and $\mathbb S^R_3$ symmetries to the left-
and right-handed fermions. Then the fermion mass matrix in a natural basis looks like
\begin{equation}
  M_f
=
  \frac {M_0} 3
\left\lgroup
\begin{matrix}
  1 & 1 & 1 \\
  1 & 1 & 1 \\
  1 & 1 & 1
\end{matrix}
\right\rgroup \,,
\label{eq:Mf0}
\end{equation}
where $M_0$ characterizes the mass scale \cite{FTY,TWY,Harari:1978yi,Koide:1989ds,Tanimoto:1989qh,Kaus:1990ij,Fritzsch:1989qm,Fritzsch:1994yx,Branco:1995pw,Fritzsch:1995dj,Xing:1996hi,Mondragon:1998gy,Fritzsch:1998xs,Fritzsch:1999ee,Haba:2000rf,Branco:2001hn,Fujii:2002jw,Fritzsch:2004xc,Rodejohann:2004qh,Teshima:2005bk,Xing:2010iu,Zhou:2011nu,Dev:2012ns,Canales:2012dr,Jora:2012nw,Yang:2016esx,Xu:2016arj,Fritzsch:2017tyf,Si:2017pdo}.
Its diagonalization involves
two mixing matrices for the left- and right-handed fermions,
$M_f = V_L D_f V^\dagger_R$ where $D_f = \mbox{diag}\{m_1, m_2, m_3\}$ is the
diagonalized mass matrix and $V_L$ ($V_R$) is the mixing matrix of left-handed
(right-handed) fermions.

The democratic mass matrix form \geqn{eq:Mf0} applies for all fermions, except
the neutrinos. This can be naturally realized with $SO(3)_L \times SO(3)_R$
flavor symmetries \cite{TWY}.
The three generations of left- and right-handed fermions form
triplets under $SO(3)_L$ and $SO(3)_R$ transformations, respectively. Similarly
there are two triplet flavons $\phi_L$ and $\phi_R$, correspondingly. Then,
two invariants can be written down to form a Yukawa term
\begin{equation}
  \sum_{ij} y_{ij}
  \left( \overline \psi_{L,i} \phi_{L,i} \right)
  \left( \phi_{R,j} \psi_{R,j} \right) \,.
\end{equation}
The $SO(3)_L \times SO(3)_R$ flavor symmetry would break down to the residual
$\mathbb S^L_3 \times \mathbb S^R_3$ if the triplet Higgs obtains equal vacuum
expectation values for the three components,
$\langle \phi_{L,R} \rangle \propto (1,1,1)$, leading to the democratic
mass matrix \geqn{eq:Mf0} for charged fermions. For neutrinos, its mass
term can be given by Weinberg-Yanagida operator \cite{Weinberg,Yanagida}
which contains two left-handed fermions, $L_{L,i} L_{L,j}$. Note that
these two fermions belong to the same $SO(3)_L$ triplet whose product
decomposes as ${\bf 3} \times {\bf 3} = {\bf 1} + {\bf 3} + {\bf 5}$.
Of the three decomposed representations, the triplet ${\bf 3}$ has
anti-symmetric Clebsch-Gordan coefficients which is not consistent with
the Majorana neutrino mass matrix. The other two, ${\bf 1}$ and ${\bf 5}$,
can give symmetric Majorana neutrino mass matrix. The singlet contribution
is proportional to a unit matrix and for the ${\bf 5}$ representation we
adopt two scalar multiplets with
\begin{equation}
  \Sigma^{(1)}_L
\propto
\left\lgroup
\begin{matrix}
  1 & 0 & 0 \\
  0 & 1 & 0 \\
  0 & 0 & -2
\end{matrix}
\right\rgroup \,,
\qquad
  \Sigma^{(2)}_L
\propto
\left\lgroup
\begin{matrix}
  1 & 0 & 0 \\
  0 &-1 & 0 \\
  0 & 0 & 0
\end{matrix}
\right\rgroup \,,
\end{equation}
to give diagonal neutrino mass matrix. Altogether, the ${\bf 1}$ and ${\bf 5}$
representations give a diagonal neutrino mass matrix with the three mass eigenvalues
being free.

The concrete form of $V_L$ is determined by
$M_f M^\dagger_f = V_L D^2_f V^\dagger_L$. Note that $M_f M^\dagger_f$ also
takes the same form as \geqn{eq:Mf0}, but with $M_0$ replaced by $M^2_0$.
By diagonalizing $M_f M^\dagger_f$, we can obtain the mixing matrix $V_L$
\begin{equation}
  V^\dagger_L
=
\left\lgroup
\begin{matrix}
  e^{i \alpha_1} \\
& e^{i \alpha_2} \\
&&e^{i \alpha_3}
\end{matrix}
\right\rgroup
\left\lgroup
\begin{matrix}
  c_T & s_T e^{i \phi} & 0 \\
- s_T e^{- i \phi} & c_T & 0 \\
  0   & 0   & 1
\end{matrix}
\right\rgroup
\left\lgroup
\begin{matrix}
  \frac {-1}{\sqrt 2} & \frac 1 {\sqrt 2}   & 0 \\
  \frac {-1}{\sqrt 6} & \frac {-1}{\sqrt 6} & \frac 2 {\sqrt 6} \\
  \frac 1 {\sqrt 3}   & \frac 1 {\sqrt 3}   & \frac 1 {\sqrt 3}
\end{matrix}
\right\rgroup
\equiv
  R T V_0 \,,
\label{eq:VL}
\end{equation}
which is the most general form of the solution. For convenience, we denote
the mixing angle among the states of degenerated mass eigenvalues $m_1 = m_2 = 0$ as
$(c_T, s_T) \equiv (\cos \theta_T, \sin \theta_T)$, to which a Dirac-type
CP phase $\phi$ is attached. In addition, there are three free rephasing
degrees of freedom $\alpha_i$ with $i = 1,2,3$ that are attached to the three
mass eigenvalues. It is interesting to observe that {\it the democratic mass
matrix leads to hierarchical mass eigenvalues.}

Since the same democratic mass matrix applies to both up- and down quarks,
the CKM matrix naturally has suppressed 1-3 and 2-3 mixings,
\begin{equation}
  V_{\rm CKM}
=
  T_u T^\dagger_d
=
\left\lgroup
\begin{matrix}
  c_{T,u} & s_{T,u} e^{i \phi_u} & 0 \\
- s_{T,u} e^{- i \phi_u} & c_{T,u} & 0 \\
  0   & 0   & 1
\end{matrix}
\right\rgroup
\left\lgroup
\begin{matrix}
  c_{T,d} & - s_{T,d} e^{i \phi_d} & 0 \\
  s_{T,d} e^{- i \phi_d} & c_{T,d} & 0 \\
  0   & 0   & 1
\end{matrix}
\right\rgroup \,.
\label{eq:CKM0}
\end{equation}
For cleanness, we have omitted the two rephasing matrices $R_u$ and $R_d$.
The combined 1-2 mixing takes the form as
\begin{equation}
  \cos \theta_{12}
=
\left|
  c_{T,u} c_{T,d}
+ s_{T,u} s_{T,d} e^{i (\phi_u - \phi_d)}
\right| 
\quad \mbox{and} \quad
  \sin \theta_{12}
=
\left|
  c_{T,u} s_{T,d}
- s_{T,u} c_{T,d} e^{i (\phi_u - \phi_d)}
\right| \,,
\end{equation}
while $\theta_{13} = \theta_{23} = 0$. This naturally explains why the mixing in
the quark sectors are small.

For the neutrino sector, \geqn{eq:VL} is already the form for the PMNS matrix,
\begin{equation}
  V_{\rm PMNS}
=
\left\lgroup
\begin{matrix}
- \frac {c_{T,\ell}} {\sqrt 2} - \frac {s_{T,\ell} e^{i \phi_\ell}}{\sqrt 6}
& \frac {c_{T,\ell}} {\sqrt 2} - \frac {s_{T,\ell} e^{i \phi_\ell}}{\sqrt 6}
& \frac {2 s_{T,\ell} e^{i \phi_\ell}}{\sqrt 6}
\\
  \frac {s_{T,\ell} e^{- i \phi_\ell}}{\sqrt 2} - \frac {c_{T,\ell}}{\sqrt 6}
&-\frac {s_{T,\ell} e^{- i \phi_\ell}}{\sqrt 2} - \frac {c_{T,\ell}}{\sqrt 6}
& \frac {2 c_{T,\ell}}{\sqrt 6}
\\
  \frac 1 {\sqrt 3} & \frac 1 {\sqrt 3} & \frac 1 {\sqrt 3}
\end{matrix}
\right\rgroup \,,
\end{equation}
assuming the neutrino mass matrix is diagonal, i.e., $V_\ell = I$.
Comparing with the standard parametrization of the PMNS matrix we can obtain
\begin{equation}
  \sin \theta_r
=
  \frac {2 s_{T,\ell}}{\sqrt 6} \,,
\qquad
  \tan \theta_a
=
  \sqrt 2 c_{T,\ell} \,,
\qquad
  \tan \theta_s
=
\left|
  \frac {\sqrt 3 c_{T,\ell} - s_{T,\ell} e^{i \phi_\ell}}
        {\sqrt 3 c_{T,\ell} + s_{T,\ell} e^{i \phi_\ell}}
\right| \,,
\qquad
  \delta_D = - \phi_\ell \,.
\label{eq:nu-sol}
\end{equation}
The mixing angles have been denoted as $(\theta_a, \theta_r, \theta_s)
\equiv (\theta_{23}, \theta_{13}, \theta_{12})$ for the atmospheric, reactor,
and solar angles, respectively, to make their physical meaning explicit.
Since there are two model parameters but four physical quantities in \geqn{eq:nu-sol},
two correlations (also called as sum rules) would emerge,
\begin{equation}
  t^2_a
=
  2 - 3 s^2_r \,,
\qquad
  \cos \delta_D
=
  \frac {1 - t^2_s}{1 + t^2_s}
  \frac {c^2_r}{s_r \sqrt{2 - 3 s^2_r}} \,.
\label{eq:sumrules}
\end{equation}
Essentially the atmospheric angle $\theta_a$ and the Dirac CP phase $\delta_D$
can be predicted as functions of the reactor and solar angles, $\theta_r$ and
$\theta_s$, which is in the same spirit as residual symmetries
\cite{Dicus:2010yu, Ge:2011ih, Ge:2011qn}.

Due to the degenerate mass eigenvalues $m_1 = m_2 = 0$,
the Dirac CP phase $\delta_D$ and the mixing $\theta_{12}$ have no physical consequence.
Nevertheless, we can still obtain some insight by playing these sum rules with
the experimentally measured values as a preliminary exercise. Since the reactor angle is small,
$s_r \approx 1/6$,  the atmospheric angle naturally lives in the second octant,
$\theta_a \approx 54^\circ$. On the other hand, the fact that
$\cos \delta_D$ cannot actually exceed 1 puts a natural limit on the
solar mixing angle
\begin{equation}
  \cos \delta_D \leq 1
\qquad \Rightarrow \qquad
  t^2_s
\geq
  \frac {c^2_r - s_r \sqrt{2 - 3 s^2_r}}
        {c^2_r + s_r \sqrt{2 - 3 s^2_r}}
\approx
  1 \,,
\end{equation}
with $s_r \ll c_r$.
Namely, the solar angle $\theta_s$ would also reside in the second octant to
allow a physical Dirac CP phase and a small reactor angle, which
is not consistent with our current knowledge on the neutrino mixing angles.
It is necessary to introduce deviation to the democratic mass matrix
\geqn{eq:Mf0} as we will discuss in the following section. Deviation
is also necessary for eliminating the two vanishing mass eigenvalues,
$m_1 = m_2 = 0$.

\section{Random Perturbations}
\label{sec:random}

As we argued above, to account for a realistic mass hierarchy with nonzero
mass eigenvalues and the measured mixing patterns, the democratic mass matrix can
only be approximate one. We would effectively break the residual
$\mathbb S^L_3$ and $\mathbb S^R_3$ symmetries. Since these two $\mathbb S_3$
symmetries are already residual ones, there is no reason to expect anything
that can still regulate the deviations. A natural scheme is that these
deviations are totally random. In addition, the deviations should not be
too far away from the zero's order which can also fit the measured neutrino
and quark mixings quite well. Consequently, the deviations should be small random
perturbations.

Generally, the mass matrix \geqn{eq:Mf0} becomes
\begin{equation}
  M_f
=
  \frac {M_0} 3
\left\lgroup
\begin{matrix}
  1 + \epsilon_{11} e^{i \phi_{11}} & 1 + \epsilon_{12} e^{i \phi_{12}} & 1 + \epsilon_{13} e^{i \phi_{13}} \\
  1 + \epsilon_{21} e^{i \phi_{21}} & 1 + \epsilon_{22} e^{i \phi_{22}} & 1 + \epsilon_{23} e^{i \phi_{23}} \\
  1 + \epsilon_{31} e^{i \phi_{31}} & 1 + \epsilon_{32} e^{i \phi_{32}} & 1 + \epsilon_{33} e^{i \phi_{33}}
\end{matrix}
\right\rgroup \,,
\label{eq:Mf}
\end{equation}
by introducing a complex perturbation $\epsilon_{ij} e^{i \phi_{ij}}$ to the
$(ij)$ element of the mass matrix, $M_{f,ij} = \frac {M_0} 3 + \delta M_{ij}$.
The flat measure of the deviations looks like $d^2 M_{ij} = d M^{(r)}_{ij} d M^{(i)}_{ij} \propto \epsilon_{ij} d \epsilon_{ij} d \phi_{ij}$ where $M^{(r)}_{ij}$ and
$M^{(i)}_{ij}$ are the real and imaginary parts of the deviations, respectively.
We assign $0 \leq \epsilon_{ij} \leq \epsilon_{max}$ and $0 \leq \phi_{ij} \leq 2 \pi$
for a random scattering. 
\begin{figure}[h]
\centering
\includegraphics[height=0.32\textwidth,width=5.25cm,angle=-90]{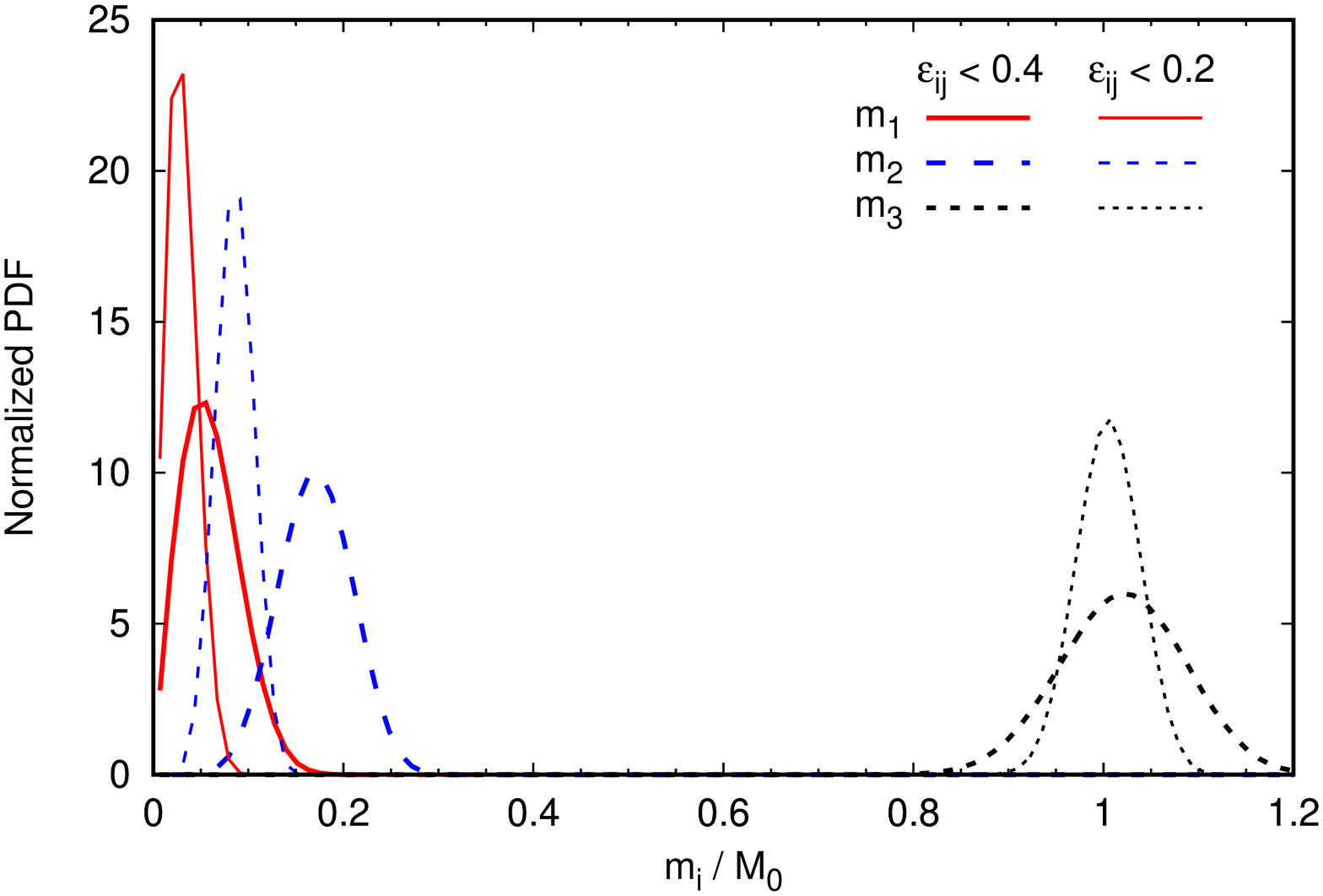}
\includegraphics[height=0.32\textwidth,width=5cm,angle=-90]{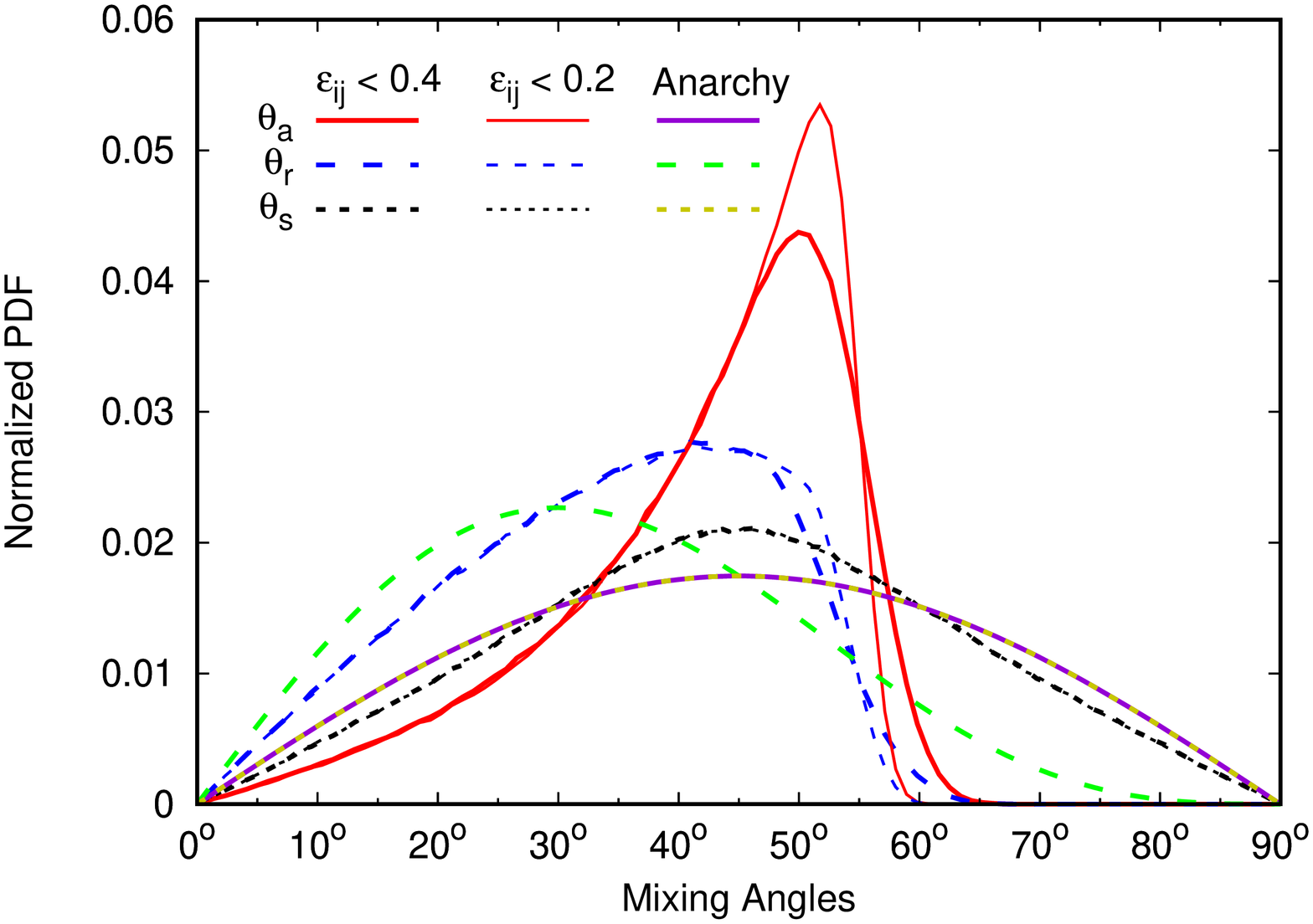}
\includegraphics[height=0.32\textwidth,width=5.25cm,angle=-90]{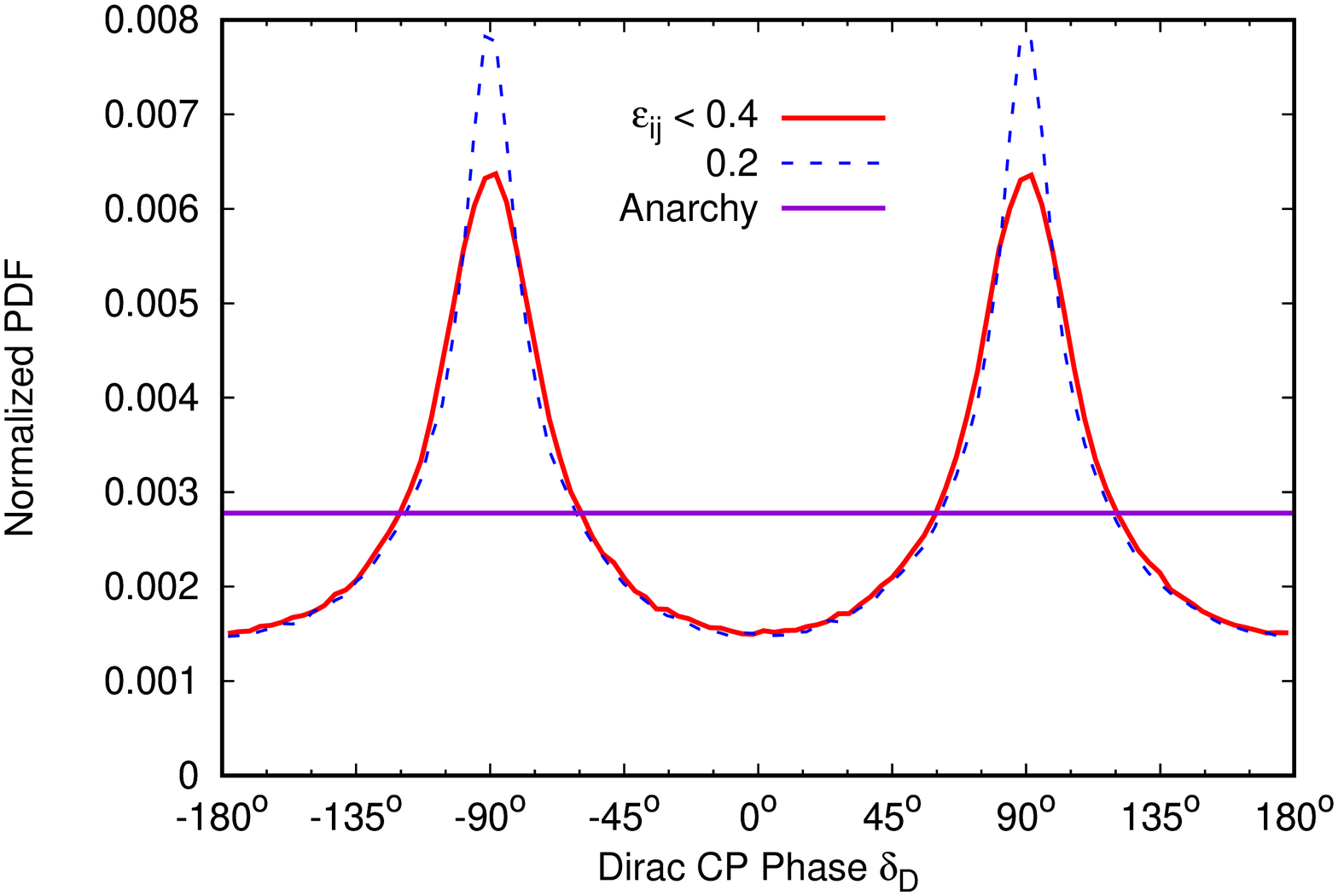}
\caption{The prediction of the charged lepton masses (Left), the neutrino mixing angles
(Middle), and the leptonic Dirac CP phase $\delta_D$ (Right) by random
perturbations to the democratic mass matrix. While the phases $\phi_{ij}$
are randomly sampled in the whole range $[0, 2 \pi]$, the deviation magnitudes are
randomly distributed in the range $0 \leq \epsilon_{ij} \leq 0.2$
(medium) or $0 \leq \epsilon_{ij} \leq 0.4$ (thick). For comparison, the prediction
of mixing angles and the Dirac CP phase from the anarchy
mass matrix is shown by thick curves with different colors.}
\label{fig:random}
\end{figure}
In \gfig{fig:random}, we show the predicted neutrino
mixing for $\epsilon_{max} = 0.2$ (medium) and 0.4 (thick).
For comparison, we also
show the prediction of mixing angles and the Dirac CP phase from the
anarchy mass matrix \cite{Haba:2000be} as thick curves with different colors.

The quasi democratic matrix can naturally explain the hierarchical structure in
fermion masses. At the leading order, the three mass eigenvalues are
$m_1 = m_2 = 0$ and $m_3 = M_0$. Introducing random perturbations can
break the degeneracy between $m_1$ and $m_2$ and at the same time broaden the
$m_3$ peak. The larger $\epsilon_{max}$, the broader mass peaks. In \gfig{fig:random}
we show predictions for both $\epsilon_{max} = 0.2$ and 0.4. An even larger
$\epsilon_{max}$ would lead to more diverse mass eigenvalues and hence more
unrealistic mass hierarchy.
For comparison,
the prediction of the anarchy mass matrix scheme is modulated by the mass eigenvalues
themselves,
$(m^2_1 - m^2_2)^2 (m^2_2 - m^2_3)^2 (m^2_3 - m^2_1)^2 m_1 m_2 m_3 dm_1 dm_2 dm_3$,
which is symmetric under the interchange of any two mass eigenvalues. In other
words, there is no preference for the mass hierarchy in the anarchy mass matrix.
The quasi democratic mass matrix leads to a much more realistic prediction of the mass hierarchy.

It is interesting to
observe that the predictions for the mixing angles and the Dirac CP phase are
quite stable against changing the perturbation size. In other words, our prediction
of the CP phases are not that sensitive to the cut-off we set on the perturbation
size. This stability feature has already been observed and quantified in the anarchy
model \cite{Haba:2000be}.

The predicted mixing angles have wide distributions. However, the predictions
of the three mixing angles clearly have different features. The atmospheric angle
$\theta_a$ tends to peak around the maximal value, $\theta_a \approx 45^\circ$
while its tail is highly suppressed on both sides, naturally explaining the measured
large value of it. In comparison, the reactor angle $\theta_r$ tends to have a
fat tail for small values, which is in accordance with its measured value
$\theta_r \approx 8.4^\circ$ \cite{deSalas:2017kay}. Both $\theta_a$ and $\theta_r$
have asymmetric distributions and an upper bound around $60^\circ$. On the other
hand, the solar angle $\theta_s$ has a symmetric probability distribution with peak
at long tail on both sides. Hence, we would expect a natural hierarchy among the
mixing angles, $\theta_r \lesssim \theta_s \lesssim \theta_a$, if all three are
in the first octant. This overall picture is quite different from the prediction of
the anarchy model where $\theta_a$ and $\theta_s$ have exactly same symmetric
distribution while $\theta_r$ has an asymmetric distribution with peak in the first
octant. The democratic mass matrix with random perturbations can
provide a better prediction than the anarchy model \cite{Haba:2000be}.

Most notably, introducing random perturbations to the democratic mass matrix
produces two prominent peaks in the probability distribution of the Dirac CP
phase around maximal values, $\delta_D \approx \pm 90^\circ$, one of which is in
perfect agreement with the current global fit \cite{deSalas:2017kay}.
In comparison, the anarchy model has no preference for any value of $\delta_D$
which appears as a flat curve in the right panel of \gfig{fig:random}.
\begin{figure}[h]
\centering
\includegraphics[width=0.8\textwidth]{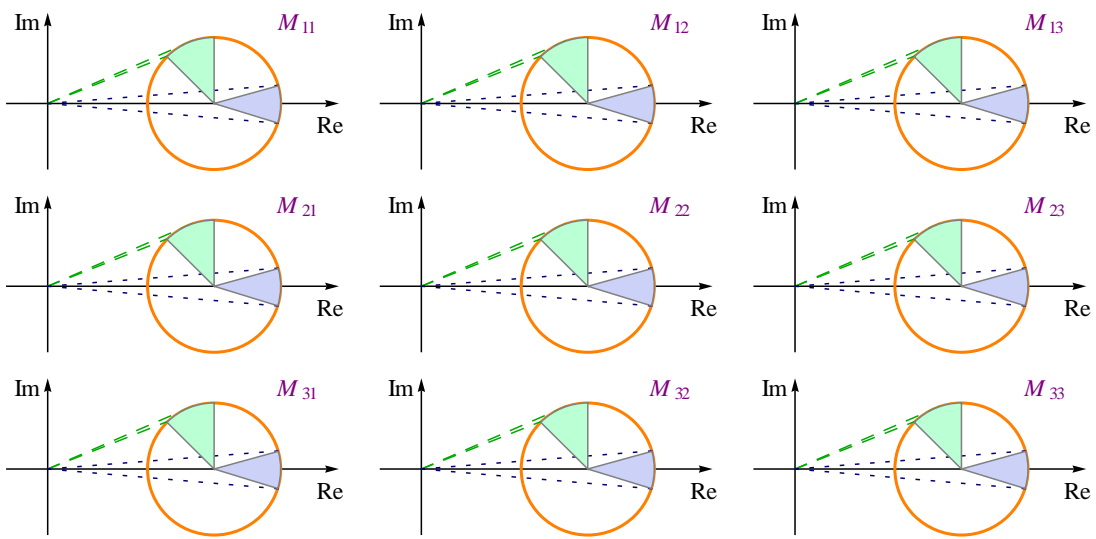}
\caption{Illustrative plot of the random perturbations with the blue patches
denoting the region with small CP violation and the green patches with large CP
violation in each perturbative deviations.}
\label{fig:Mij}
\end{figure}

We use the illustrative \gfig{fig:Mij} to show how these two CP peaks appear as
a natural consequence of the random perturbations. For each element of the
fermion mass matrix, there are two regions with small CP violation (one of which
labeled as blue sector) and two regions with large CP violation (one of which
labeled as green sector). For simplicity, we consider only these two extreme cases.
Suppose the probability for small and large CP violation in each element is $P_s$
and $P_l$, respectively. Since we have omitted the intermediate regions, there is
no need to require probability conservation, $P_s + P_l \leq 1$.
The physical Dirac CP phase $\delta_D$ comes from the interplay of
CP phases in all matrix elements. If a random large CP appears in every element
of the mass matrix, it is very difficult for the physical Dirac CP phase to be
small which can only happen when all the CP phases $\phi_{ij}$ and the matrix
element size $\epsilon_{ij}$ are highly fine tuned to cancel each other.
Consequently, a small Dirac CP phase usually appears when all the matrix elements
have small CP. The probability for small $\delta_D$ is of the order $P^9_s$.
In contrast, it is enough to have large Dirac CP phase if one of the matrix element
has a large CP. So the probability for large $\delta_D$ can be roughly estimated
as $9 P_l$. The factor that $P_s < 1$ naturally leads to $P^9_s \ll 9 P_l$, in other
words, there is much larger probability for maximal Dirac CP phase than the vanishing one.
\begin{figure}[h]
\centering
\includegraphics[height=0.6\textwidth,angle=-90]{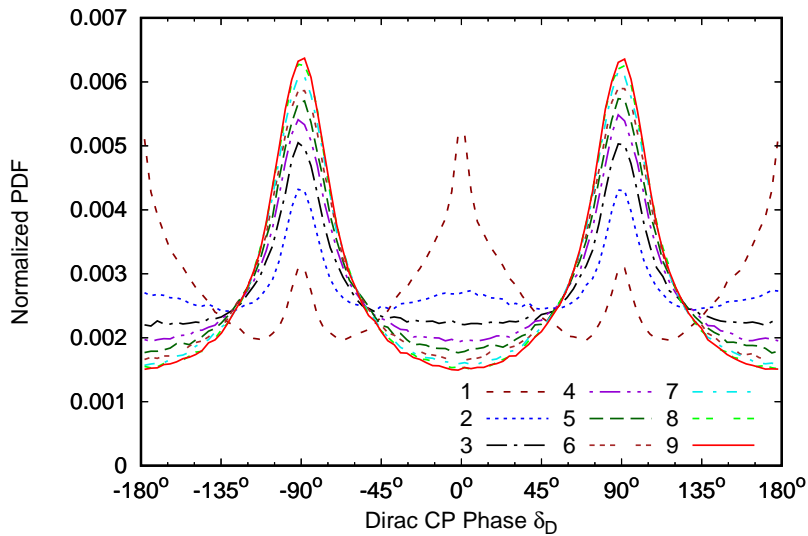}
\caption{The predicted CP phase distributions with randomly selecting which
input phase $\phi_{ij}$ can be nonzero and take randomly sampled values.
For the curve labeled with $n$, $n$ of the 9 input phases are selected.}
\label{fig:phases}
\end{figure}
 
To make it explicit, we show predictions of the Dirac CP phase
distributions with randomly selected input phases $\phi_{ij}$ in \gfig{fig:phases}.
For each curve, we randomly select $n$ of the 9 input phases. The value of the
selected phase $\phi_{ij}$ is randomly sampled in the range $[0, 2 \pi]$ while
those unselected are set to be zero. We can clearly see the trend that with
more input CP phases randomly sampled, the Dirac CP phase is easier to have
large value.

\section{Comparison with Measurements}
\label{sec:comparison}

\begin{figure}[h]
\centering
\includegraphics[height=0.32\textwidth,width=5.25cm,angle=-90]{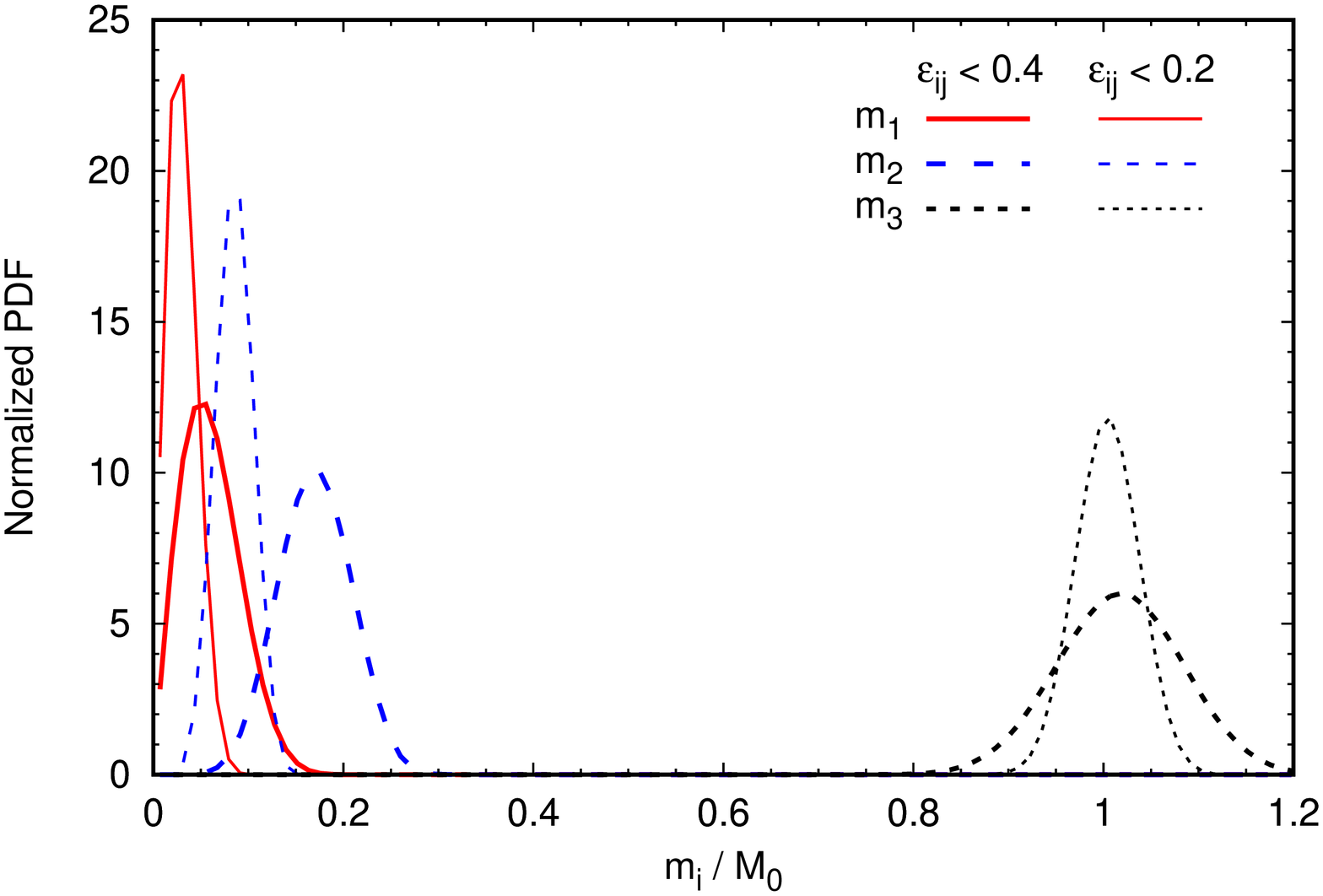}
\includegraphics[height=0.32\textwidth,width=5cm,angle=-90]{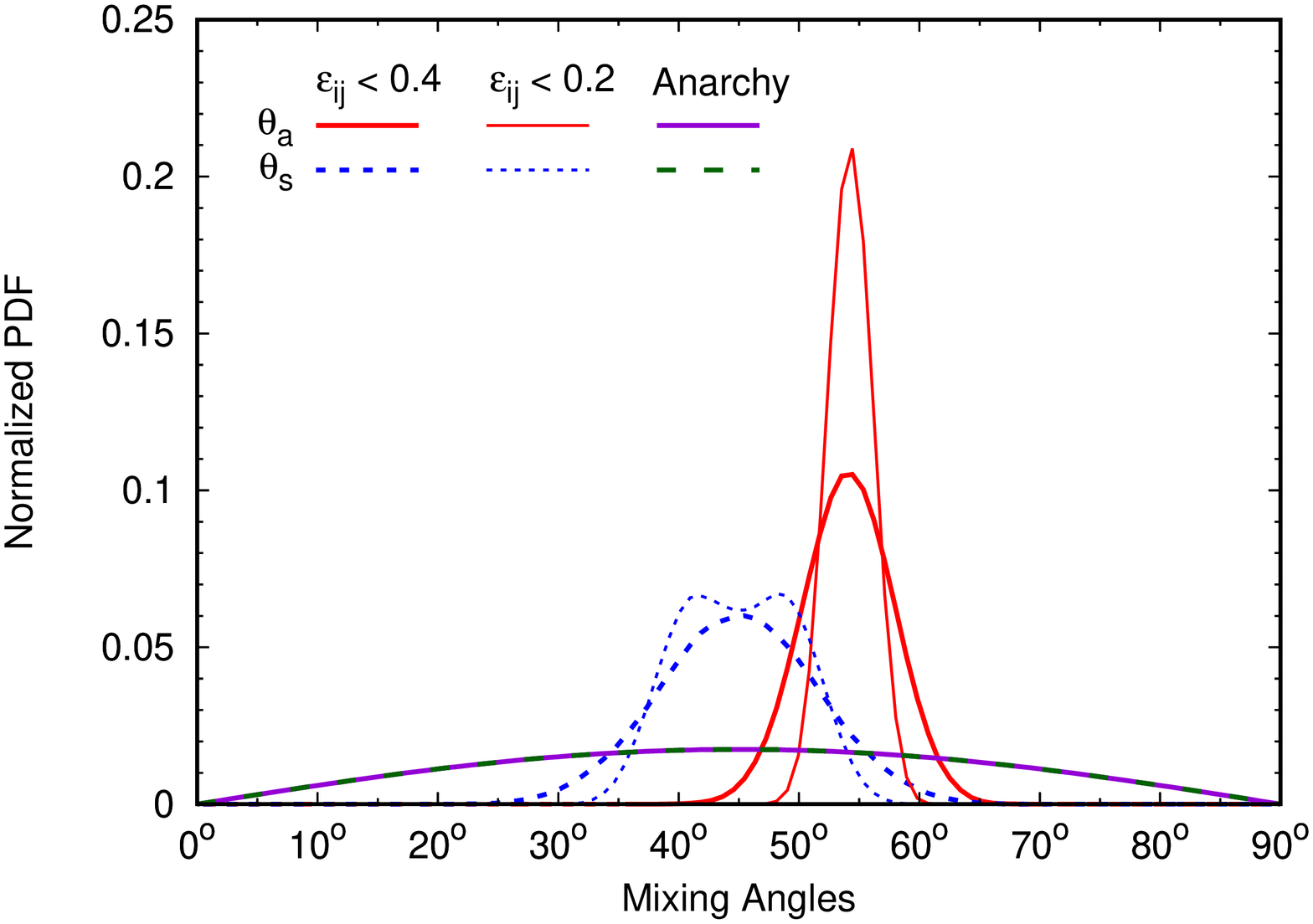}
\includegraphics[height=0.32\textwidth,width=5.25cm,angle=-90]{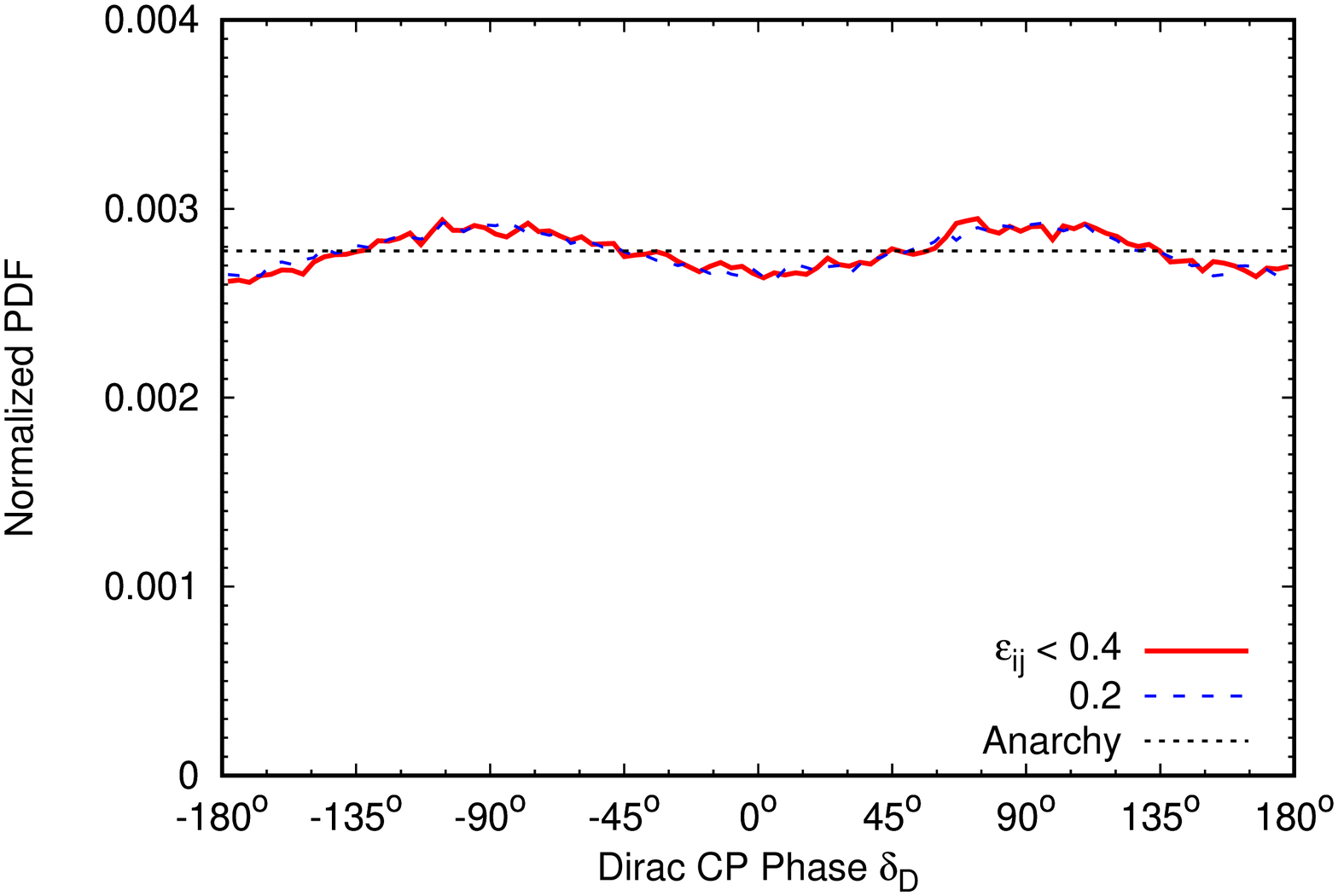}
\caption{The prediction of the charged lepton masses (Left), the neutrino mixing angles
(Middle), and the leptonic Dirac CP phase $\delta_D$ (Right) by random 
perturbations of the democratic mass matrix. We show randomly sampled
phases $0 \leq \phi_{ij} \leq 2 \pi$ and magnitude $0 \leq \epsilon_{ij} \leq 0.2$
(medium line) or $0 \leq \epsilon_{ij} \leq 0.4$ (thick line), while the reactor angle $\theta_r$
is constrained to be in the $3 \sigma$ range of the current global fit,
$8.44^\circ \pm (3 \times 0.16)^\circ$. For comparison, the prediction
of mixing angles and the Dirac CP phase from the anarchy model for complex
mass matrix, $d s^2_s d c^4_r d s^2_a d \delta_D$, is shown as thick curves with different colors.}
\label{fig:selectTr}
\end{figure}

In principle, the prediction of our model should come from unbiased sampling
of the mass matrix element as we have shown in the previous section with even distributions according
to the flat measure $\epsilon_{ij} d \epsilon_{ij} d \phi_{ij}$. If the flavor
mixing pattern is really determined by dice, the values of the mixing parameters
are determined once for all and our current knowledge from experimental measurements
would not affect their probability distributions. Anyhow, to make the features
of our predictions more explicit, we show some interesting features by constraining
the most precisely measured reactor angle $\theta_r$, as shown in \gfig{fig:selectTr}.

The most significant effect of constraining the reactor angle $\theta_r$ appears in
the predictions of mixing angles and the Dirac CP phase. In particular, the
distributions of $\theta_a$ and $\theta_s$ shrink a lot and hence they are more predictive.
As consistent to the zeroth order sum rule \geqn{eq:sumrules}, the atmospheric angle
$\theta_a$ now resides in the second octant which can be readily tested by future
measurements. For the solar angle $\theta_s$, although it is still
symmetric around the maximal value, its distribution also significantly shrinks
while the true value $\theta_s \approx 34.5^\circ$ is still covered with sizable
probability. On the other hand, the CP probability only has weak preference for
maximal CP violation now. The tendency of having large $\cos \delta_D$
for a small reactor angle $\sin \theta_r \approx 1/6$ as demonstrated by
\geqn{eq:sumrules} cancels the preference of large CP induced by the democratic
mass matrix with random perturbations as explained in the previous section. However, the CP phase
distribution in \gfig{fig:selectTr} is not worse than the anarchy model
prediction. 

\section{Quark Mixing}
\label{sec:quark}

\begin{figure}[h]
\centering
\includegraphics[height=0.4\textwidth,angle=-90]{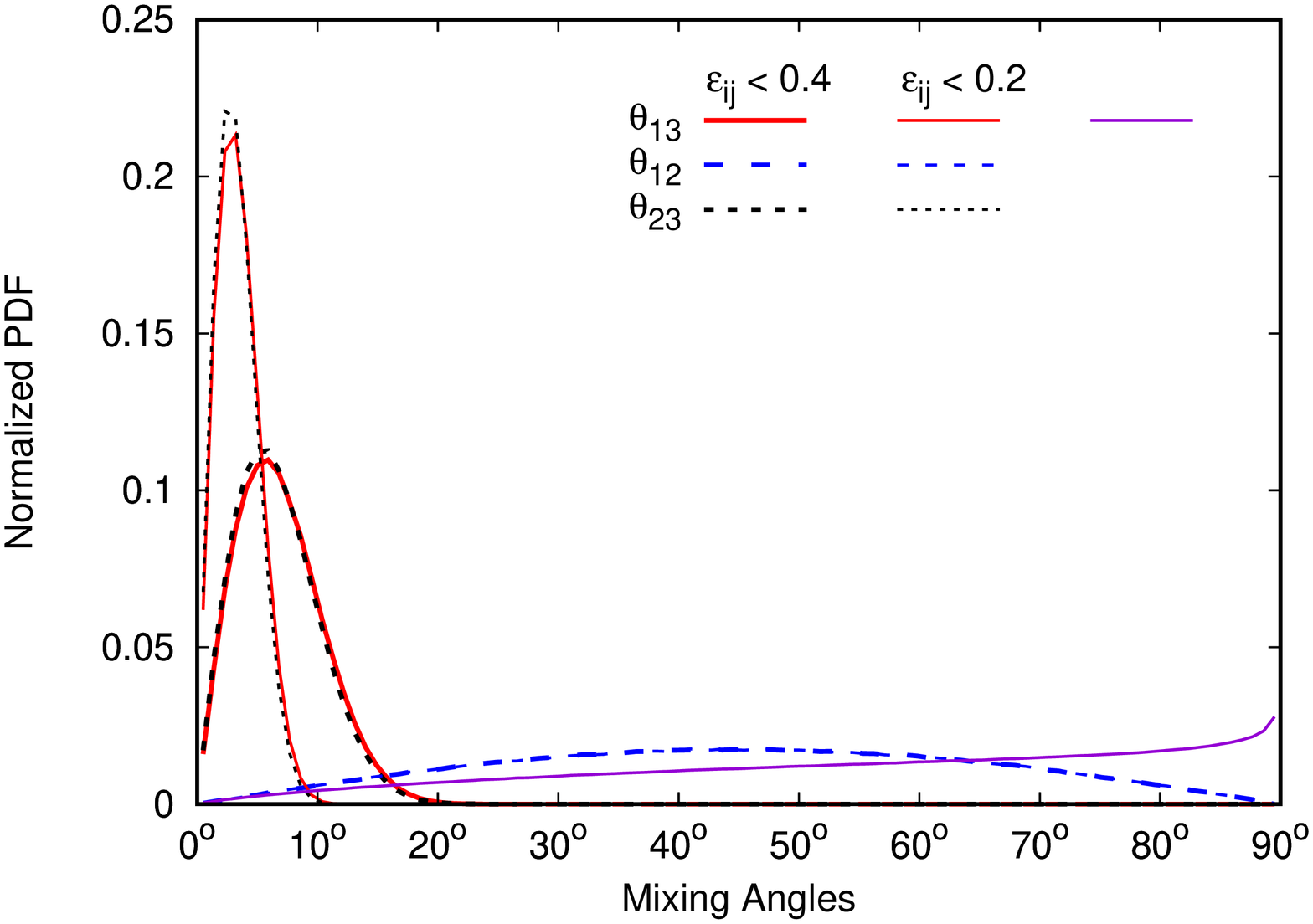}
\includegraphics[height=0.4\textwidth,angle=-90]{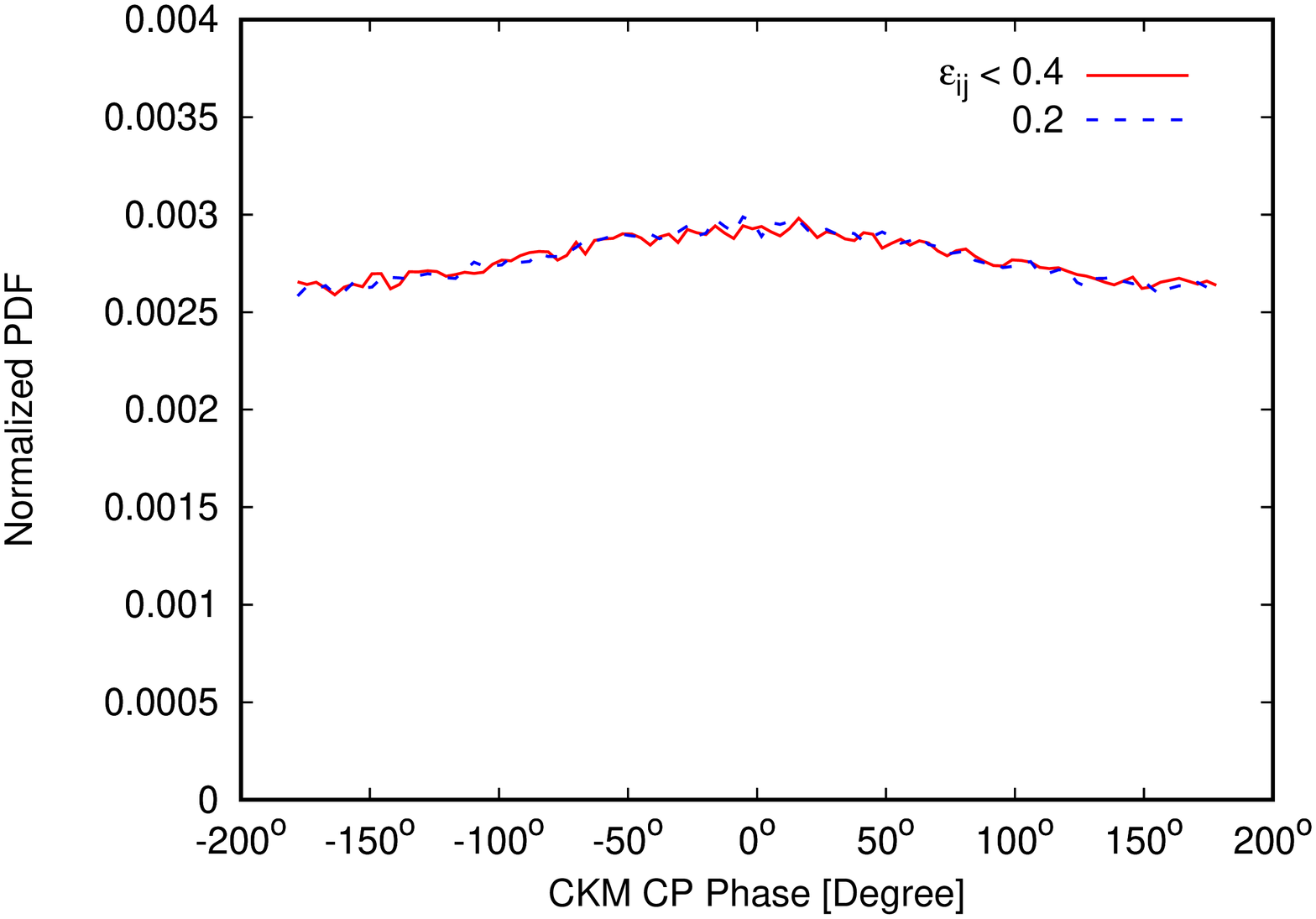}
\caption{The prediction of the quark mixing angles (Left) and the quark CP
phase (Right) by random perturbations to the democratic mass matrix
for both up and down quarks. While the phases $\phi_{ij}$ of
perturbations are randomly sampled in the whole range $[0, 2 \pi]$,
the deviation size are randomly distributed in the range $0 \leq \epsilon_{ij} \leq 0.2$
(medium) or $0 \leq \epsilon_{ij} \leq 0.4$ (thick). For comparison, the prediction
of $\theta_{12}$ at leading order is shown
as a solid purple line in the left panel.}
\label{fig:quark}
\end{figure}

The democratic mass matrix with random perturbations can not only explain the
neutrino mixing but also the quark mixing. As demonstrated in \gsec{sec:M0},
the quark mixing has naturally suppressed $\theta_{13}$ and $\theta_{23}$ while
the $\theta_{12}$ can take any value, see \geqn{eq:CKM0}. To make it exact,
the measure of the $T_u$ and $T_d$ transformations are
$d T_u d T_d = d \theta_{T,u} d \phi_u d \theta_{T,d} d \phi_d$.
By randomly sampling $\theta_{T,u}$, $\theta_{T,u}$, and $\phi_u - \phi_d$,
we obtain a distribution of $\theta_{12}$, shown as a purple solid line in
\gfig{fig:quark}. At leading order, the $\theta_{12}$ tends to have large
value with peak at $90^\circ$. Note that this is just a preliminary result
as explained in \gsec{sec:M0} and the prediction can be improved by random perturbations
as we discuss below.

Introducing random perturbations to the democratic mass matrix
\geqn{eq:Mf0} can naturally produce nonzero $\theta_{13}$ and $\theta_{23}$
as shown in \gfig{fig:quark}. Since there is degeneracy between the two
lightest mass eigenvalues, namely there is no difference between them,
$\theta_{13}$ and $\theta_{23}$ have exactly the same probability distribution.
The spread of of $\theta_{13}$ and $\theta_{23}$ distributions is closely
related to the perturbation size. This is closely related to the quark mass
distribution which takes exactly the same form as the charged lepton mass
distribution in \gfig{fig:random}. In other words, the hierarchical mixing
in the quark sector is closely related to the quark mass hierarchy. On the
other hand, $\theta_{12}$ has wide symmetric distribution around the
maximal value $45^\circ$ which is already significantly improved. In addition, the distribution of $\theta_{12}$
is actually independent of $\theta_{13}$ and $\theta_{23}$ as indicated by
the zeroth-order analysis in \gsec{sec:M0}, in contrast to the lepton case.
For the CKM CP phase, its distribution is almost flat with weak preference
of zero value. Although sampling multiple input CP phases in a single mass
matrix can enhance the probability of large physical CP phase, the combined
prediction from two mass matrices cancels any preference.

For comparison, the anarchy model prediction of quark mixing,
$U_{\rm CKM} = U^\dagger_u U_d$, takes exactly the same form as the
anarchy model prediction of neutrino mixing,
$U_{\rm PMNS} = U^\dagger_\ell$, shown in the middle panel of \gfig{fig:random}
as thick curves with different colors.
This is because left and right rotations can not affect the
Haar measure, $d U_{\rm CKM} = d U_d = d U^\dagger_u$. Since all Dirac fermions
are subject to a $3 \times 3$ complex mass matrix with the same $U(3)$ Haar
measure for their mixing, $d U_d = d U^\dagger_u = d U^\ell_\ell$, and hence
$d U_{\rm CKM} = d U_{\rm PMNS}$, which we have also checked by numerically sampling
the up and down quark mixings. The democratic mass matrix with
random perturbations gives a much better prediction than the anarchy model,
especially for $\theta_{13}$ and $\theta_{23}$.

For quark mass matrices, the predictions of broken democracy can depend on the energy scale at which the symmetry is broken, because renormalization group running of couplings can alter and amplify the symmetry breaking effects.  This is in contrast with the case of neutrino masses and mixings, where the effects of running are small due to the smallness if the symmetry breaking Yukawa couplings.

\section{Conclusions}

We have considered a broken democracy model with
residual $\mathbb S_3$ symmetries to dictate democratic mass 
matrices for both up and down quarks
as well as for charged leptons.
This naturally explains
why the CKM matrix has only a sizable 1-2 mixing while the PMNS 
matrix can has two large mixing angles. In addition, this 
assignment also leads to massless quarks and charged 
leptons in the first two generations. To account for the
measured values of neutrino mixing angles and
the nonzero fermion masses, the residual $\mathbb S_3$
symmetries have to be broken. Since the residual symmetry is 
already the one that survives symmetry breaking by definition, 
there is no other fundamental principle but random
deviations to regulate the mass matrices after the
residual $\mathbb S_3$ symmetries are broken. With
the general features fixed by residual symmetries at
leading order, the random deviations can only be perturbative.
Our broken democracy model with residual $\mathbb S_3$
naturally leads to a large leptonic Dirac CP phase with two
peaks around $\pm \pi/2$, in addition to the realistic mixing
patterns and mass hierarchy in both quark and lepton sectors.

We have assumed in this paper that the neutrino mass matrix is diagonal ($V_\ell = I$). The neutrino masses are given by the Weinberg-Yanagida operator \cite{Weinberg,Yanagida}, $(\overline{L^c_\alpha} \widetilde H^*) (\widetilde H^\dagger L_\beta) / M$ where $L_\alpha$ are lepton doublets and $\widetilde H \equiv \epsilon H^*$ is the CP conjugation of the Higgs doublet $H$, in the standard model, which gives us two $\mathbb S^L_3$-invatiant mass matrices such as
\begin{equation}
\left\lgroup
\begin{matrix}
  a & 0 & 0 \\
  0 & a & 0 \\
  0 & 0 & a
\end{matrix}
\right\rgroup
\qquad \mbox{and} \qquad
\left\lgroup
\begin{matrix}
  0 & b & b \\
  b & 0 & b \\
  b & b & 0
\end{matrix}
\right\rgroup \,.
\end{equation}
However, even if we choose the first matrix for the neutrinos, we have a problem. This is because this choice predicts the degenerate masses for the neutrinos which is strongly disfavored by the observations. Therefore, we need a large violation of the $\mathbb S^L_3$ symmetry. As shown in \gsec{sec:M0}, this problem may be solved in a model based on the $SO(3)_L \times SO(3)_R$ symmetry \cite{TWY}.

\section{Acknowledgements}
We thank Hitoshi Murayama for encouragement of this work.  This work was supported by World Premier International Research Center Initiative (WPI), MEXT, Japan. S.F.G. was also supported by JSPS KAKENHI Grant Number JP18K13536. A.K. was also supported by the U.S. Department of Energy Grant No. DE-SC0009937.
T.T.Y. was also supported by Grants-in-Aid for Scientific Research
from the Ministry of Education, Culture, Sports, Science, and Technology (MEXT) KAKENHI, Japan, No. 26104001, No. 26104009, No. 16H02176, and No. 17H02878.


\end{document}